\documentclass{article}[16pt]
\usepackage[utf8]{inputenc}
\usepackage{graphicx}
\usepackage{lscape}
\begin{document}
\begin{center}

{\bf \Large GRADIENTS OF METALLICITY AND AGE OF STARS IN THE DWARF SPHEROIDAL GALAXIES KKs~3 AND ESO~269-66}

\bigskip

{\bf M. E. Sharina, L. N. Makarova, and D. I. Makarov}

\bigskip

{\bf Special Astrophysical Obseratory of the Russian Academy of Sciences, Nizhnyi Arkhiz, Russia\\ e-mail: sme@sao.ru, lidia@sao.ru, dim@sao.ru}
  \end{center}
  
\bigskip

{\em Abstract. 
We compare the properties of the stellar populations of the globular clusters and field stars in two dwarf
spheroidal galaxies (dSphs): ESO~269-66, a near neighbor of the giant S0 galaxy NGC 5128, and KKs~3,
one of the few extremely isolated dSphs within 10 Mpc. The histories of star formation in these galaxies
are known from previous work on deep stellar photometry using images from the Hubble Space Telescope
(HST). The age and metal content for the nuclear globular clusters in KKs~3 and ESO~269-66 are known
from literature spectroscopic studies: $T=12.6$~billion years, $[Fe/H]=-1.5$ and -1.55~dex. 
 We use the
Sersic function to construct and analyze the profiles of the surface density of the stars with high and low
metallicities (red and blue) in KKs~3 and ESO~269-66, and show that (1) the profiles of the density of red
stars are steeper than those of blue stars, which is indicative of gradients of metallicity and age in the
galaxies, and (2) the globular clusters in KKs~3 and ESO~269-66 contain roughly 4 and 40\%, respectively, 
of all the old stars in the galaxies with metallicities $[Fe/H]\sim-1.5\div-1.6$~dex and ages of 12-14 billion years.
 The globular clusters are, therefore, relicts of the first, most powerful bursts of star
formation in the central regions of these objects. It is probable that, because of its isolation, KKs~3 has
lost a smaller fraction of old low-metallicity stars than ESO~269-66.} \\

\bigskip

{\bf 1. Introduction}

\bigskip
 
Dwarf spheroidal (dSphs) and elliptical (dEs) galaxies contain no stars younger than a billion years. As
opposed to dEs, dSphs have a low surface brightness $SB_{0,B} > 22.5-23$ mag arcsec$^{-2}$ (in the $B$ filter of the 
Johnson-Cousins photometric system). 
As opposed to irregular dwarf galaxies with a low surface brightness, spheroidal galaxies are
not detected in the neutral hydrogen (HI) lines and do not contain active star formation regions. With a few exceptions,
almost all dSphs are observed within ~300~kpc  around massive galaxies [1-4].
This observation is characterized by
the ''morphology-density'' relation [5-7]. The physical mechanisms responsible for this effect include tidal interactions 
of galaxies, ram pressure stripping of gas in the intergalactic medium in groups and clusters of
galaxies, and ejection of gas by stellar wind from supernova outbursts [2,8-13]. The more massive galaxies can absorb
less massive ones during close transits.
 
 Gradients of the properties of the stellar populations exist in the dSphs of the local group [14-15] and dE
satellites of M31: NGC 205, 185, and 147 [16-17]. Observations show that the average stellar metallicity is higher
in the central regions of the galaxies than at their peripheries and the average age is younger. The existence of the
gradients explains the evolutionary behavior of these objects. The magnitude of the gradients can be explained if
it is assumed that the dEs and dSphs were low-mass spirals in the past [18]. It is possible that the destroyed spirals,
which have lost gas and some of their stars, may be the result of collisions of two or more dwarf galaxies [19].
 
KKs~3 and ESO~269-66 exist in different environments, but have similar star formation histories with three
main periods, of which the most ancient, 10-12 Gyr ago, was the most intense [4,20,21]. KKs~3 is unique, a very
isolated dSph within 10 Mpc [4]. ESO~269-66 is a near neighbor of the giant S0 NGC 5128 [22]. This dSph is unusual
in the large dispersion of the metallicity of its stars [22,23]. The properties of these galaxies are listed in 
Table~\ref{tab:properties}  with the relevant references: equatorial coordinates, color excess, distance to the objects, 
their diameters, color in the international broad-band Johnson-Cousins system, absolute V band stellar magnitude, 
approximate mass of HI, and average metallicity.

\begin{table}
\caption{Observed Properties of KKs~3 and ESO~269-66.}
\begin{center}
\begin{tabular}{l|cc|cl}
\hline\hline
                                              &  KKs\,3                                      & Ref. &  ESO\,269-66                                &  Ref.  \\   \hline\hline
RA(J200.0)                                    & $2^h24^m44^s4$                           & [4]  & $\rm 13^h13^m09^s1$                     &  [25]   \\
Dec.(J200.0)                                  & $\rm -73^o30^{\prime\prime}51^{\prime}$  & [4]  & $\rm -44^o53^{\prime\prime}24^{\prime}$ &  [25]   \\
$\rm E(B-V)$                                  & 0.045                                    & [4]  & 0.093                                  &  [25]   \\ 
Distance, Mpc                                 & 2.12                                     & [4]  & 3.82                                    &  [25]    \\
Diameter, kpc                                 & 1.5                                     & [4]  & 2.4                                     &  [25]    \\
$\rm (V-I)$, mag                              & 0.77                                    & [24] & 1.06                                    &  [23]    \\
$\rm M_V$                                     & -12.3                                    & [24] & -14.4                                   &  [25]    \\
 $\rm M_{HI}$ , $M_{\odot}$                   & 1.1$\cdot 10^{5}$                        & [24] & $<0.9\cdot 10^{5}$                     &  [25]     \\
$[Fe/H]_{12 \div14Gyr}$,dex                   & -1.9                                     &  [4] & -1.75                                   &  [20]      \\
\hline\hline                                                                                                                                 
\end{tabular}
\end{center}
\label{tab:properties}
\end{table}

\bigskip

{\bf 2. Star-formation bursts in KKs~3 and ESO~269-66 and the formation of globular clusters}

\bigskip

The histories of star formation in KKs~3 and ESO~269-66 (studied in Refs. 4 and 20) are summarized in Table~\ref{tab:SFH_KKs3}. 
 where the first column lists the metallicity in dex, the second, the time interval in billions of years, the third, the
level (range) of star formation in $M_{\odot}/yr$, and the fourth, the resulting mass of the stars in $M_{\odot}$. The table shows
that there were several star formation outbursts in both of the galaxies. Despite the different number of outbursts
and different masses of the stars, several common features can be noticed: (1) the first outbursts, which took place
within about 2 Gyr of the formation of the universe, were the most powerful; (2) they were followed by less intense
events over a period of 2-6 Gyr ago; and, finally, (3) the final star-formation periods, which occurred about 1 Gyr
ago, swept the gaseous residues from the galaxies. At present, no young star-formation regions are observed in these
objects. Based on models of the star-formation histories, the average metallicities and ranges of metallicity in stars
with different ages observed in KKs~3 and ESO~269-66 are proportional to the masses of the galaxies, i.e., to the
amount of gas involved in the star-formation process. In ESO~269-66 the average metallicity of the stars is higher
and the range of metallicity is larger: $[Fe/H] =-1.8 \div 0.2$~dex for ESO~269-66 and $[Fe/H] =-2.4 \div -0.7$~dex for KKs~3.

\begin{table}
\caption{ Details of the Star-Formation Histories of KKs~3 [4] and ESO
269-66 [20].}
\vspace{3mm}
\begin{center}
\begin{tabular}{lcll}
\hline\hline
$[Fe/H]$      &  $T$           & $SFR$                      & $ M_{stars}$  \\
\hline\hline  
              &                &   KKs~3                    &               \\ \hline                                                 
 -2.36        &  [ 12 - 14 ]   & [ 0.34e-03$\div$4.47e-03 ] & 4.12e+06  \\
 -1.74        &  [ 12 - 14 ]   & [ 0.35e-03$\div$1.37e-02 ] & 1.33e+07    \\
 -1.33        &   [ 4 - 6 ]    & [ 0.14e-03$\div$3.39e-03 ] & 3.25e+06    \\
 -0.72        &   [ 0.5 - 1 ]  & [ 0.11e-03$\div$2.83e-03 ] & 7.37e+05     \\
 -0.72        & [ 1.5 - 2 ]    & [ 0.21e-03$\div$8.05e-03 ] &  1.96e+06    \\ \hline
               &                & ESO~269-66                 &              \\ \hline                                          
 -1.74        &  [ 12 - 14 ]   & [ 2.55e-03$\div$10.35e-03 ]& 6.50e+06  \\
 -1.33        &  [ 12 - 14 ]   & [ 3.00e-03$\div$2.08e-01 ] & 1.74e+08    \\
 -0.72        &   [ 1.5 - 2 ]  & [ 7.88e-04$\div$4.43e-03 ] & 8.65e+05    \\
 -0.72        &   [ 2 - 4 ]    & [ 6.90e-04$\div$9.79e-03 ] & 9.07e+06     \\
 -0.72        &   [ 4 - 6 ]    & [ 1.24e-03$\div$9.45e-03 ] &  8.17e+06    \\
  -0.41        & [ 1.5 - 2 ]   & [ 2.70e-03$\div$1.12e-01 ] &  2.73e+07    \\
  0.18         & [ 0 - 0.5 ]   & [ 1.30e-03$\div$6.70e-02 ] &  1.64e+07    \\
\hline\hline
\end{tabular}
\end{center}
\label{tab:SFH_KKs3}
\end{table}
\begin{figure}[hbt]
\begin{center}
\includegraphics[scale=0.35,angle=-90]{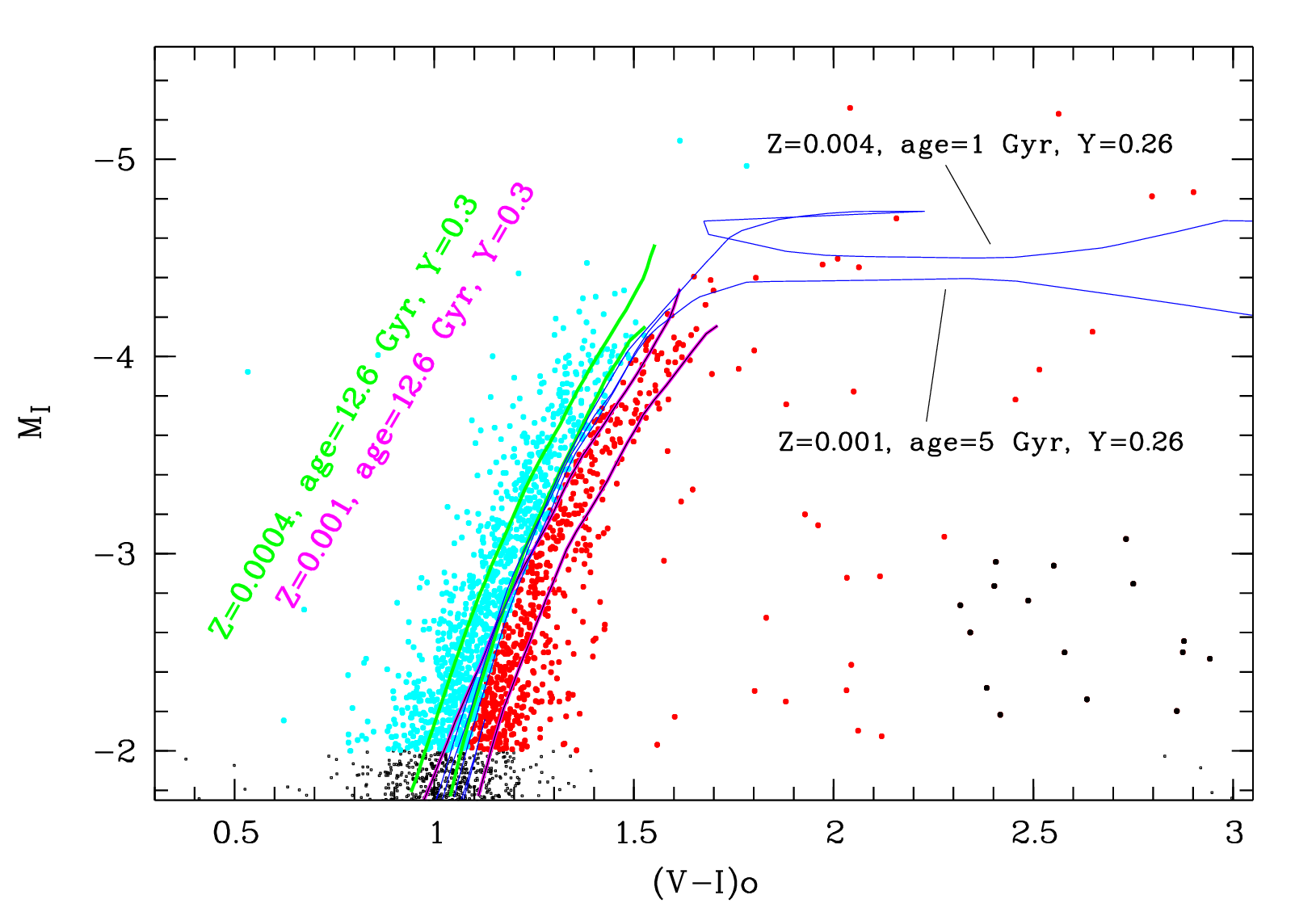}
\includegraphics[scale=0.35,angle=-90]{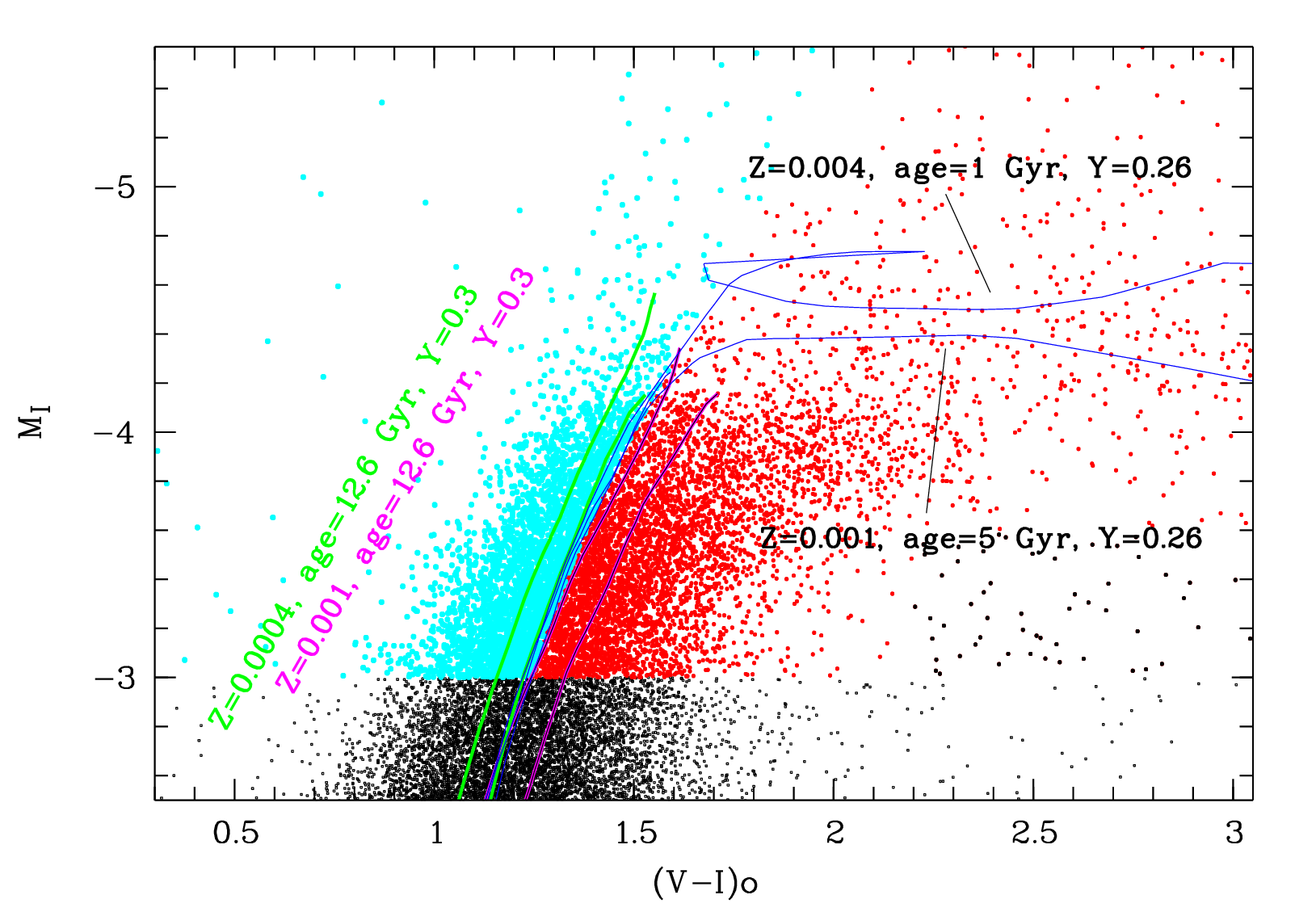}
\caption{"Color-magnitude diagrams" for KKs~3 (a) and ESO~269-66 (b). The blue stars (blue points) and red stars
(red points) were selected as shown in the figures.}
\label{CMDselect}
\end{center}
\end{figure}
\begin{figure}[hbt]
\begin{center}
\includegraphics[scale=0.35,angle=-90]{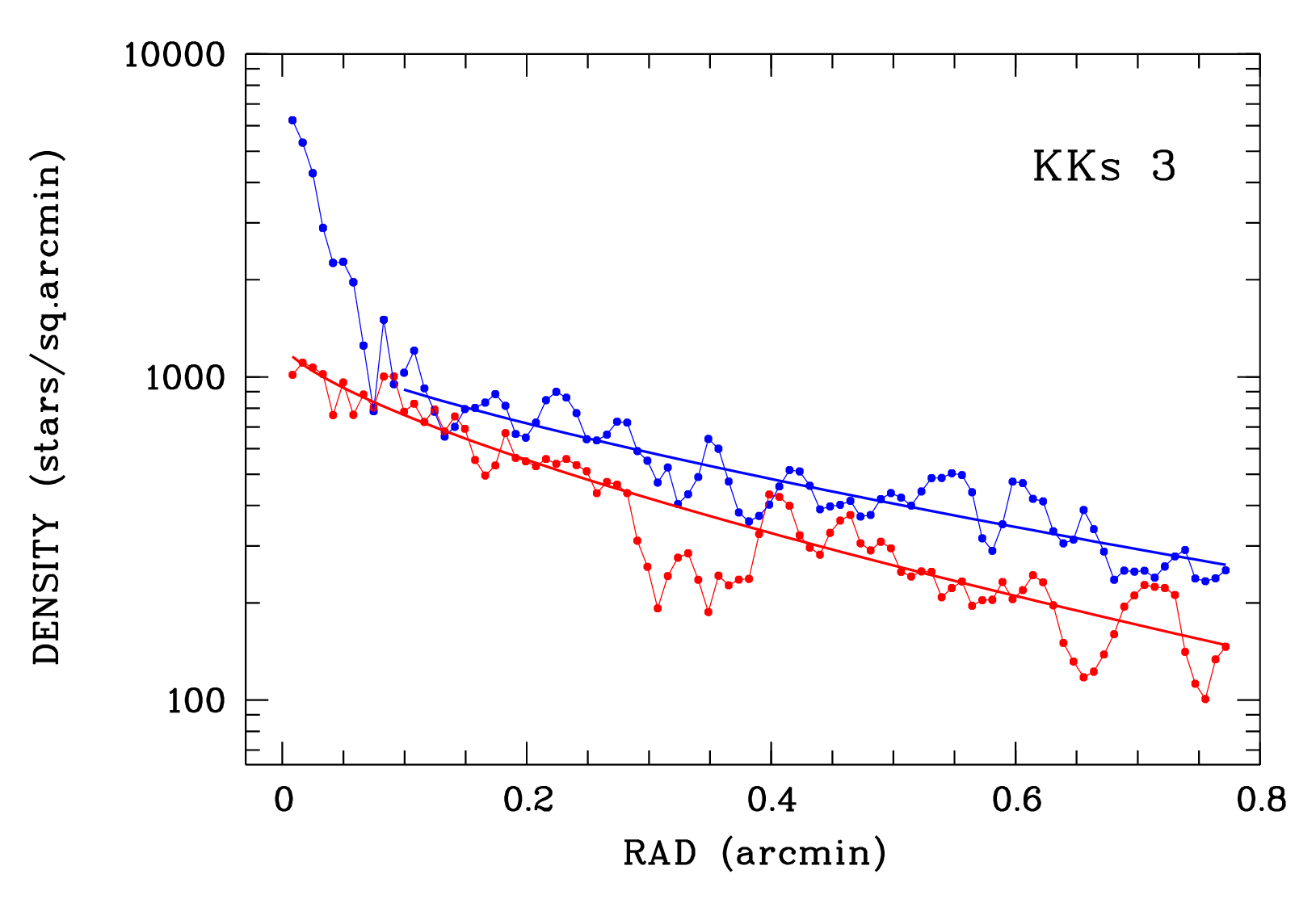}
\includegraphics[scale=0.35,angle=-90]{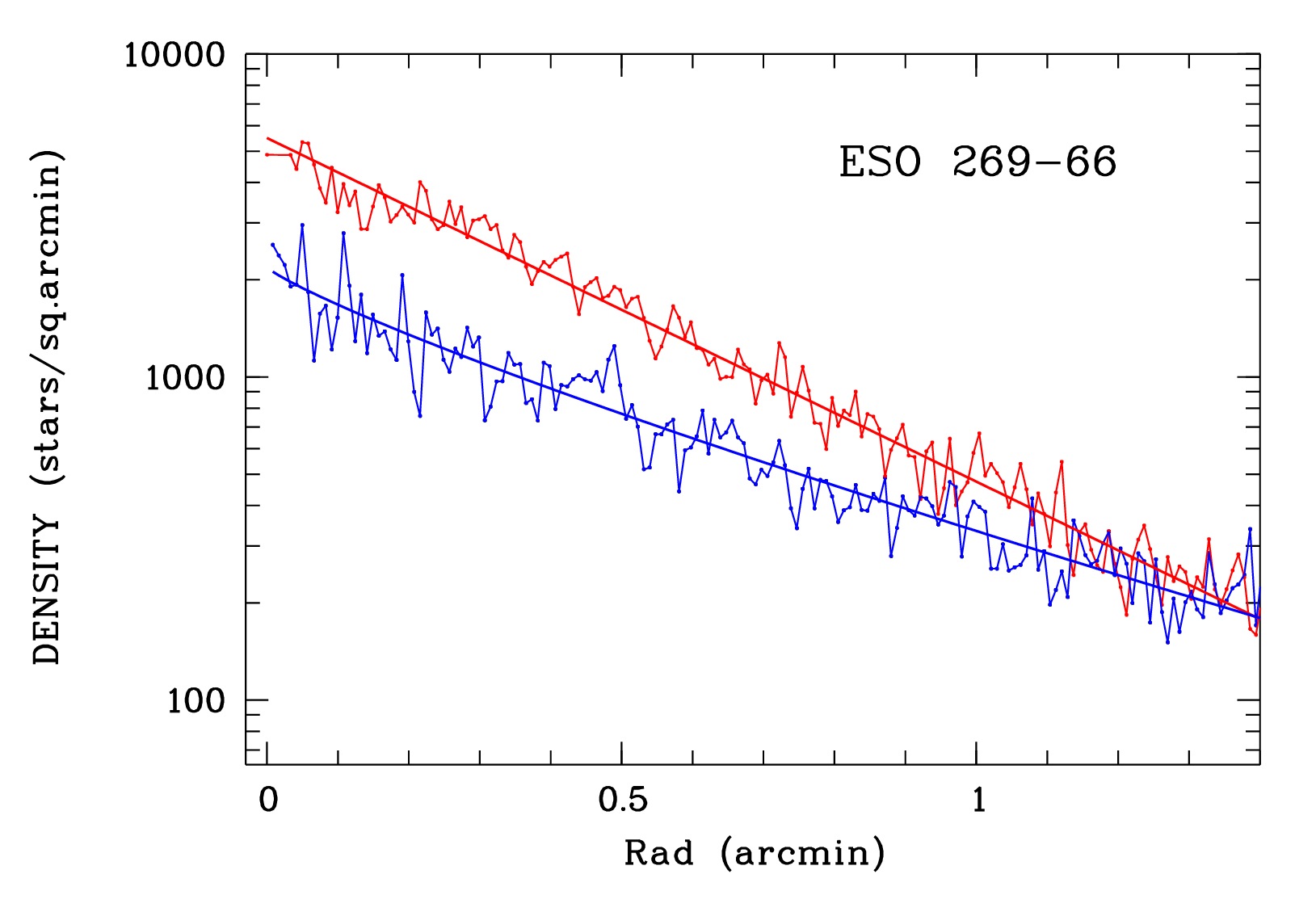}
\caption{Surface density profiles for blue (dark grey) and red (light grey) stars in KKs~3 (a) and ESO~269-66 (b) with
superimposed Sersic model curves. The contribution of the galactic background (~1 star/sq.arcmin) has been subtracted
from each point in the profiles. }
\label{DensityRad}
\end{center}
\end{figure}

The age and metallicity of the central globular clusters in  KKs~3 and ESO~269-66, 
$T=12.61\pm1$~Gyr, $[Fe/H]=-1.5, -1.55\pm0.2$~dex [26]], lie within intervals corresponding to the two earliest periods of star formation
in these galaxies. The data of Table~\ref{tab:SFH_KKs3},
can be used to calculate the fraction of stars with metallicity $[Fe/H]\sim-1.6$~dex
dex contained in the globular clusters. Roughly 57 and 27\%, respectively, of the stars in KKs~3 and ESO~269-66 have $[Fe/H]\sim-1.6$~dex.
 The absolute magnitudes of the globular clusters in KKs~3 and ESO~269-66 are $M_v = -8.48^m$ and $M_v = -9.9^m$, respectively. 
Taking the average mass to luminosity ratio of the stars in the globular clusters to
be $M/L=3$,  the masses of the globular clusters in KKs~3 and ESO~269-66 are $6.33e+05 M_{\odot}$ and $2.34e+06 M_{\odot}$, respectively.
 Thus, the globular clusters contain roughly 4 and 40\%, respectively, of all the stars in KKs~3 and ESO~269-66 
 with ages of 12-14 Gyr and $[Fe/H]\sim-1.6$~dex. Based on the surrounding dwarf galaxies, we can assume that
the isolated KKs~3 lost a fraction of stars with ages on the order of the age of the Universe that was 10 times smaller
than that lost by ESO~269-66, which lies near the giant S0 NGC~5128 in the central region of the group of Centaurus~A.

\bigskip

{\bf 2. The distributions of red and blue galaxies in KKs~3 and ESO~269-66.}

It has been shown [26] that the stars within radii of
 $\sim$47~pc  around the centers of the globular clusters in KKs~3 and ESO~269-66 have systematically more bluish color 
 (lower metallicities) than the average for the stars in the rest of these galaxies. 
 The situation is very similar in the two galaxies, despite the different distances to them and
the fact that the width of the red giant branches (RGB) owing to the spread in the stars with respect to metallicity
and age is substantially different for KKs~3 and ESO~269-66. It should be noted that in the HST images of KKs~3
and ESO~269-66, only stars in the RGB and the asymptotic giant branches (AGB) can be seen. At a level of
 $M_I \geq -3.5\pm0.1$  close to 100\% of these stars are detected in these HST images according to ''artificial star''
experiments [4,20]. In ESO~269-66 the dispersion $(V-I)$ in the colors of the stars at a level of $M_I = -3.5\pm0.1$ 
 for the I absolute stellar magnitude in the I-band of the Johnson-Cousins system is $\sigma(V-I)=0.19^m$ [23].
For KKs~3 this value is $\sigma(V-I)=0.08^m$ [26]. 

It is interesting to check how the stars in the blue and red parts of the ''color-magnitude diagram'' (CMD) are
distributed over the body of the galaxies. The method for separating the stars into blue and red is illustrated in Fig.~\ref{CMDselect}. 
The nominal separation curve is the part of the isochrone from [27] with metallicity $Z=0.0004$, an age of $age=1$~Gyr and a helium 
content of $Y=0.26$. It should be noted that the color distribution of the stars on the CMD is subject
to the so-called ''age-metallicity'' degeneracy effect. Thus, younger stars with higher metallicities can seem bluer than
stars with lower metallicity but greater age. Thus, the age and metallicity of representatives of simple stellar
populations can be judged in terms of their colors only in a probabilistic sense.

Of the 36763 stars in ESO~269-66 for which photometry was carried out and the results are shown in Fig.~\ref{CMDselect},
5753 red and 3389 blue stars with low photometric errors with a detection coverage of $\sim$100\% and $M_I < -3^m$ 
were selected for subsequent analysis.
The method for selecting the red and blue stars is illustrated in Fig.~\ref{CMDselect}.

The CMD of KKs~3 (Fig.~\ref{CMDselect}) contains 22707 stars.  Of these, 632 red and 1032 blue stars were selected by
the method shown in Fig.~\ref{CMDselect}. These are stars with low photometric errors, a detection coverage of $\sim$100\% and $M_I < -2^m$.
The sampling depth is different for the two galaxies because of the different photometric depth of the diagrams,
mainly because of the different distances from these objects.

\bigskip

{\bf 2. Density profiles of the stars}

\bigskip

To construct stellar density profiles for KKs~3 and ESO~269-66 we used the stellar photometry data and isolated
the blue and red stars with small photometric errors and a detection coverage of $\sim$100\%, from the whole data set, as
described in the previous section.
 Then we divided the two-dimensional distribution of the stars over the surface
of a galaxy into cells with sizes of 0.5 arcsec and counted the number of stars contained in each cell. After subtraction
of the background around the galaxies, the results of this count served as the basis for constructing the stellar density
profiles for KKs~3 and ESO~269-66.

The azimuthally averaged distributions of the stellar density in rings around the centers of the galaxies
coinciding with the centers of the globular clusters are shown in Fig.~\ref{DensityRad}.  It can be seen that the blue stars in KKs~3 and
ESO~269-66 (that is, the statistically older and low-metallicity stars) have flatter density profiles. The red stars are
more concentrated toward the centers of the galaxies. 

For modelling the shapes of the density profiles as a function of the radius r of the galaxies, we have used
the Sersic function [28]: $$I(r)=I_0\cdot e^{-\nu_n (\frac{r}{r_e})^{1/n}},$$
where $I_0$ is the central intensity, $r_e$ is the effective radius, 
$n$  is a positive real number representing the degree of curvature of the profile, and 
$\nu_n$ is a constant chosen so that half the total luminosity is radiated within the effective [29] radius with
$$\nu_n \sim 2n-1/3 +4/(405 n) +46/(25515 n^2).$$
 In units of surface brightness, this will be
$$ \mu(r) = \mu_0 + \frac{2.5 \nu_n}{ln10} (\frac{r}{r_e})^{1/n},$$
where $\mu_0$ is the central surface brightness.
 The parameters of the Sersic function that we have chosen for the
distributions of the red and blue stars in KKs~3 and ESO~269-66 are shown in Table~\ref{tab:Sersic}, where
$SD_{centr}$  is the central surface brightness, $r_{eff}$ is the effective radius, and $n$ is the shape factor for the profile.
  It can be seen that in accordance
with the measured parameters, the effective radii of the distributions for the blue stars are greater than those for the
red stars. We note, however, that this visually evident tendency in Fig.~\ref{DensityRad}  is less distinct and is more difficult to model
in the case of KKs~3, since the density of stars in this galaxy is much lower than in ESO~269-66. In the case of the
blue stars in KKs~3, we could only make an estimate of the parameters $r_{eff}$ and $n$ of the Sersic function. These are given in 
Table~\ref{tab:Sersic}  with colons without indication of an error.
\begin{table}
\caption{\label{tab:Sersic} Parameters of the Sersic Model [28,29]}
\begin{center}
\begin{tabular}{lccc}
\hline\hline
Object & $lg(SD_{centr})$         & $r_{eff}$      & $n$        \\
          &  [stars/arcmin$^2$]      & [arcmin]     &           \\
\hline  \hline                                                     
KKs3          &                     &              &           \\
Blue stars& $3.12\pm0.08$       &  4.82:       &  1.38:    \\
Red stars& $3.10\pm0.04$       & $3.49\pm1.13$& $1.41\pm0.16$\\ \hline
ESO~269-66    &                     &              &              \\
Blue stars& $3.34\pm0.03$       & $3.24\pm0.46$& $1.18\pm0.08$\\
Red stars& $3.74\pm0.03$       & $1.71\pm0.10$& $1.00\pm0.05$\\
\hline\hline
\end{tabular}
\end{center}
\end{table}

\bigskip

{\bf 5. Conclusion.}

\bigskip

This study of the distribution of blue and red stars in the two dSph KKs~3 and ESO~269-66, which exist in
different environments, has shown that both galaxies have gradients in their stellar populations; in particular, the older
and low-metallicity blue stars have a flatter density profile than the younger and higher metallicity red stars. Thus,
as in massive spiral galaxies, older, low-metallicity stars predominate in the outer regions of these two dwarf galaxies.
Residues of the first most powerful star formation outbursts exist in the centers of the galaxies in the form of globular
clusters which contain 4 and 40\% of the stars in KKs~3 and ESO~269-66, respectively, with metallicities on the order
of $[Fe/H]\sim-1.5\div-1.6$~dex and ages of 12-14~Gyr. 
A substantial fraction of the stars produced in the central stellar
outbursts has been lost. The globular clusters have also lost stars. As shown in Ref. 26, in the outer parts of the central globular cluster
in KKs~3 an increase is observed in the brightness of the stars within the limits of the tidal radius which represents
about 10\% of the total luminosity of the cluster. At this site the brightness profile of the cluster differs substantially
from the King's law [30]. ESO~269-66 has evidently lost a larger fraction of stars in the course of its evolution than
has the isolated KKs~3. Based on the photometric data alone, is has not been possible to describe the star formation
process in KKs~3 and ESO~269-66 in more detail and to discuss the reasons for the loss of stars. High resolution
spectroscopy of globular clusters and individual stars in dSphs in the future may aid in understanding the formation
and evolution of nuclear stellar clusters and their host galaxies.

\clearpage

\bigskip

  {\bf References}
  
\bigskip

1. Binggeli B., in Meylan G., Prugniel P., eds, ESO/OHP Workshop on warf Galaxies. ESO, Garching, 

p. 123 (1994).

2. Grebel E. K., Gallagher J. S. III, Harbeck D., Astron. J., 125, 1926 (2003).

3. Karachentsev I. D., Karachentseva V. E., Sharina M. E., in Jerjen H., Binggeli B., eds, IAU Colloq. 198, 

Near-field Cosmology with Dwarf Elliptical Galaxies. Cambridge Univ. Press, Cambridge, p. 105, (2005).

4. Karachentsev I. D., Makarova L. N., Makarov D. I., Tully R. B., Rizzi L., Mon. Not. Roy. Astron. Soc., 

447, L85 (2015).

5. Hubble, E. \& Humason, M. L., Astrophys. J., 74, 43 (1931).

6. Einasto J., Saar E., Kaasik A., Chernin A. D., Nature, 252, 111 (1974).

7. Dressler A., Astrophys. J., 236, 351 (1980).

8. Gunn J. E., Gott J. R. I., Astrophys. J., 176, 1 (1972).

9. Zasov A. V., Karachentseva V. E., Sov. Astron. Lett., 5, 137 (1979).

10. Dekel A., Silk J., Astrophys. J., 303, 39 (1986).

11. Ferrara A., Tolstoy E., Mon. Not. Roy. Astron. Soc., 313, 291 (2000).

12. Gnedin O. Y., Astrophys. J., 589, 752 (2003).

13. Boselli A., Gavazzi G., Publ. Astron. Soc. Pacific, 118, 517 (2006).

14. Harbeck D. et al., Astron. J., 122, 3092 (2001).

15. McConnachie, A. W., Astron. J., 144, 4 (2012).

16. Davidge T. J., Astron. J., 130, 2087 (2005).

17. Sharina M. E., Afanasiev V. L., Puzia T. H., Mon. Not. Roy. Astron. Soc., 372, 1259 (2006).

18. Kormendy, J., Bender, R., Astrophys. J. Suppl. Ser., 198, 2 (2012).

19. V{\"a}is{\"a}nen P., Barway S., Randriamanakoto Z., Astrophys. J. Letters, 797 (2014).

20. Makarova L., et al., Proceedings IAU Symposium 235: Galaxy Evolution across the Hubble Time, 

Eds. F.Combes and J.Palous, 320 (2007).

21. Crnojevi{\'c} D., Rejkuba M., Grebel E. K., da Costa G., Jerjen H., 2011, Astron. and Astrophys., 

530, 58 (2011).

22. Karachentsev I. D. et al., Astron. J., 133, 504 (2007).

23. Sharina M. E. et al., Mon. Not. Roy. Astron. Soc., 384, 1544 (2008).

24. Karachentsev I. D., Kniazev A. Yu., Sharina M. E., Astron. Nachr., 336, 707 (2015).

25. Karachentsev, I.D.; Makarov, D.I.; Kaisina, E.I., Astron. J., 145, 101 (2013).
                    
26. Sharina, M. E.; Shimansky, V. V.; Kniazev, A. Y., Mon. Not. Roy. Astron. Soc., 471, 1955 (2017).

27. Bertelli G., Girardi L., Marigo P., Nasi E., Astron. and Astrophys., 484, 815 (2008).

28. Sersic, J.~L., Atlas de Galaxias Australes, Cordoba, Argentina: Observatorio Astronomico (1968).

29. Ciotti, L.; Bertin, G., Astron. and Astrophys., 352, 447 (1999).

30. King I., Astron. J., 67, 471 (1962).

\end{document}